\documentclass[10pt,journal,epsfig]{IEEEtran}

\usepackage[dvips]{graphicx}
\usepackage{graphicx}
\usepackage{amssymb}
\usepackage{cite}
\usepackage{subfigure}
\usepackage{amsmath}

\begin{document}

\title{Pattern-Coupled Sparse Bayesian Learning
for Recovery of Block-Sparse Signals}


\author{Jun Fang, Yanning Shen, Hongbin Li,~\IEEEmembership{Senior
Member,~IEEE}, and Pu Wang
\thanks{Jun Fang, Yanning Shen are with the National Key Laboratory on Communications,
University of Electronic Science and Technology of China, Chengdu
611731, China, Emails: JunFang@uestc.edu.cn,
201121260110@std.uestc.edu.cn}
\thanks{Hongbin Li is with the Department of Electrical and
Computer Engineering, Stevens Institute of Technology, Hoboken, NJ
07030, USA, E-mail: Hongbin.Li@stevens.edu}
\thanks{Pu Wang is with the Schlumberger-Doll Research Center, Cambridge,
MA, 02139, USA, Email: pwang@ieee.org}
\thanks{This work was supported in part by the National
Science Foundation of China under Grant 61172114, and the National
Science Foundation under Grant ECCS-0901066.}}

\maketitle


\begin{abstract}
We consider the problem of recovering block-sparse signals whose
structures are unknown \emph{a priori}. Block-sparse signals with
nonzero coefficients occurring in clusters arise naturally in many
practical scenarios. However, the knowledge of the block structure
is usually unavailable in practice. In this paper, we develop a
new sparse Bayesian learning method for recovery of block-sparse
signals with unknown cluster patterns. Specifically, a
pattern-coupled hierarchical Gaussian prior model is introduced to
characterize the statistical dependencies among coefficients, in
which a set of hyperparameters are employed to control the
sparsity of signal coefficients. Unlike the conventional sparse
Bayesian learning framework in which each individual
hyperparameter is associated independently with each coefficient,
in this paper, the prior for each coefficient not only involves
its own hyperparameter, but also the hyperparameters of its
immediate neighbors. In doing this way, the sparsity patterns of
neighboring coefficients are related to each other and the
hierarchical model has the potential to encourage
structured-sparse solutions. The hyperparameters, along with the
sparse signal, are learned by maximizing their posterior
probability via an expectation-maximization (EM) algorithm.
Numerical results show that the proposed algorithm presents
uniform superiority over other existing methods in a series of
experiments.
\end{abstract}



\section{Introduction}
Compressive sensing is a recently emerged technique of signal
sampling and reconstruction, the main purpose of which is to
recover sparse signals from much fewer linear measurements
\cite{ChenDonoho98,CandesTao05,Donoho06}
\begin{align}
\boldsymbol{y}=\boldsymbol{Ax}
\end{align}
where $\boldsymbol{A}\in\mathbb{R}^{m\times n}$ is the sampling
matrix with $m\ll n$, and $\boldsymbol{x}$ denotes the
$n$-dimensional sparse signal with only $K$ nonzero coefficients.
Such a problem has been extensively studied and a variety of
algorithms that provide consistent recovery performance guarantee
were proposed, e.g.
\cite{ChenDonoho98,CandesTao05,Donoho06,TroppGilbert07,Wainwright09,DaiMilenkovic09}.
In practice, sparse signals usually have additional structures
that can be exploited to enhance the recovery performance. For
example, the atomic decomposition of multi-band signals
\cite{Mishali09} or audio signals \cite{GribonvalBacry03} usually
results in a block-sparse structure in which the nonzero
coefficients occur in clusters. In addition, a discrete wavelet
transform of an image naturally yields a tree structure of the
wavelet coefficients, with each wavelet coefficient serving as a
``parent'' for a few ``children'' coefficients \cite{HeCarin09}. A
number of algorithms, e.g., block-OMP \cite{EldarKuppinger10},
mixed $\ell_2/\ell_1$ norm-minimization \cite{EldarMishali09},
group LASSO \cite{YuanLin06}, StructOMP \cite{HuangZhang11}, and
model-based CoSaMP \cite{BaraniukCevher10} were proposed for
recovery of block-sparse signals, and their recovery behaviors
were analyzed in terms of the model-based restricted isometry
property (RIP) \cite{EldarMishali09,BaraniukCevher10} and the
mutual coherence \cite{EldarKuppinger10}. Analyses suggested that
exploiting the inherent structure of sparse signals helps improve
the recovery performance considerably. These algorithms, albeit
effective, require the knowledge of the block structure (such as
locations and sizes of blocks) of sparse signals \emph{a priori}.
In practice, however, the prior information about the block
structure of sparse signals is often unavailable. For example, we
know that images have structured sparse representations but the
exact tree structure of the coefficients is unknown to us. To
address this difficulty, a hierarchical Bayesian
``spike-and-slab'' prior model is introduced in
\cite{HeCarin09,YuSun12} to encourage the sparseness and promote
the cluster patterns simultaneously. Nevertheless, for both works
\cite{HeCarin09,YuSun12}, the posterior distribution cannot be
derived analytically, and a Markov chain Monte Carlo (MCMC)
sampling method has to be employed for Bayesian inference. In
\cite{PelegEldar12,DremeauHerzet12}, a graphical prior, also
referred to as the ``Boltzmann machine'', was used to model the
statistical dependencies between atoms. Specifically, the
Boltzmann machine is employed as a prior on the support of a
sparse representation. However, the maximum a posterior (MAP)
estimator with such a prior involves an exhaustive search over all
possible sparsity patterns. To overcome the intractability of the
combinatorial search, a greedy method \cite{PelegEldar12} and a
variational mean-field approximation method \cite{DremeauHerzet12}
were proposed to approximate the MAP. Recently, a sparse Bayesian
learning method was proposed in \cite{ZhangRao13} to address the
sparse signal recovery problem when the block structure is
unknown. In \cite{ZhangRao13}, the components of the signal are
partitioned into a number of overlapping blocks and each block is
assigned a Gaussian prior. An expanded model is then used to
convert the overlapping structure into a block diagonal structure
so that the conventional block sparse Bayesian learning algorithm
can be readily applied.

In this paper, we develop a new Bayesian method for block-sparse
signal recovery when the block-sparse patterns are entirely
unknown. Similar to the conventional sparse Bayesian learning
approach \cite{Tipping01,JiXue08}, a Bayesian hierarchical
Gaussian framework is employed to model the sparse prior, in which
a set of hyperparameters are introduced to characterize the
Gaussian prior and control the sparsity of the signal components.
Conventional sparse learning approaches, however, assume
independence between the elements of the sparse signal.
Specifically, each individual hyperparameter is associated
independently with each coefficient of the sparse signal. To model
the block-sparse patterns, in this paper, we propose a coupled
hierarchical Gaussian framework in which the sparsity of each
coefficient is controlled not only by its own hyperparameter, but
also by the hyperparameters of its immediate neighbors. Such a
prior encourages clustered patterns and suppresses ``isolated
coefficients'' whose pattern is different from that of its
neighboring coefficients. An expectation-maximization (EM)
algorithm is developed to learn the hyperparameters characterizing
the coupled hierarchical model and to estimate the block-sparse
signal. Our proposed algorithm not only admits a simple iterative
procedure for Bayesian inference. It also demonstrates superiority
over other existing methods for block-sparse signal recovery.



The rest of the paper is organized as follows. In Section
\ref{sec:model}, we introduce a new coupled hierarchical Gaussian
framework to model the sparse prior and the dependencies among the
signal components. An expectation-maximization (EM) algorithm is
developed in Section \ref{sec:inference} to learn the
hyperparameters characterizing the coupled hierarchical model and
to estimate the block-sparse signal. Section
\ref{sec:inference-un} extends the proposed Bayesian inference
method to the scenario where the observation noise variance is
unknown. Relation of our work to other existing works are
discussed in \ref{sec:discussions}, and an iterative reweighted
algorithm is proposed for the recovery of block-sparse signals.
Simulation results are provided in Section \ref{sec:simulation},
followed by concluding remarks in Section \ref{sec:conclusion}.





\section{Hierarchical Prior Model} \label{sec:model}
We consider the problem of recovering a block-sparse signal
$\boldsymbol{x}\in\mathbb{R}^{n}$ from noise-corrupted
measurements
\begin{align}
\boldsymbol{y}=\boldsymbol{A}\boldsymbol{x}+\boldsymbol{w}
\end{align}
where $\boldsymbol{A}\in\mathbb{R}^{m\times n}$ ($m<n$) is the
measurement matrix, and $\boldsymbol{w}$ is the additive
multivariate Gaussian noise with zero mean and covariance matrix
$\sigma^2\boldsymbol{I}$. The signal $\boldsymbol{x}$ has a
block-sparse structure but the exact block pattern such as the
location and size of each block is unavailable to us.

In the conventional sparse Bayesian learning framework, to
encourage the sparsity of the estimated signal, $\boldsymbol{x}$
is assigned a Gaussian prior distribution
\begin{align}
p(\boldsymbol{x}|\boldsymbol{\alpha})=\prod_{i=1}^n
p(x_i|\alpha_i)
\end{align}
where $p(x_i|\alpha_i)=\mathcal{N}(x_i|0,\alpha_i^{-1})$, and
$\boldsymbol{\alpha}\triangleq\{\alpha_i\}$ are non-negative
hyperparameters controlling the sparsity of the signal
$\boldsymbol{x}$. Clearly, when $\alpha_i$ approaches infinity,
the corresponding coefficient $x_i$ becomes zero. By placing
hyperpriors over $\{\alpha_i\}$, the hyperparameters
$\{\alpha_i\}$ can be learned by maximizing their posterior
probability. We see that in the above conventional hierarchical
Bayesian model, each hyperparameter is associated independently
with each coefficient. The prior model assumes independence among
coefficients and has no potential to encourage clustered sparse
solutions.





%

To exploit the statistical dependencies among coefficients, we
propose a new hierarchical Bayesian model in which the prior for
each coefficient not only involves its own hyperparameter, but
also the hyperparameters of its immediate neighbors. Specifically,
a prior over $\boldsymbol{x}$ is given by
\begin{align}
p(\boldsymbol{x}|\boldsymbol{\alpha})=&\prod_{i=1}^n
p(x_i|\alpha_i,\alpha_{i+1},\alpha_{i-1}) \label{eq3}
\end{align}
where
\begin{align}
p(x_i|\alpha_i,\alpha_{i+1},\alpha_{i-1})=
\mathcal{N}(x_i|0,(\alpha_i+\beta\alpha_{i+1}+\beta\alpha_{i-1})^{-1})
\label{prior}
\end{align}
and we assume $\alpha_0=0$ and $\alpha_{n+1}=0$ for the end points
$x_1$ and $x_n$, $0\leq \beta\leq 1$ is a parameter indicating the
relevance between the coefficient $x_i$ and its neighboring
coefficients $\{x_{i+1},x_{i-1}\}$. To better understand this
prior model, we can rewrite (\ref{prior}) as
\begin{align}
p(x_i|\alpha_i,\alpha_{i+1},\alpha_{i-1})\propto
p(x_i|\alpha_i)[p(x_i|\alpha_{i+1})]^{\beta}[p(x_i|\alpha_{i-1})]^{\beta}
\label{eq2}
\end{align}
where $p(x_i|\alpha_j)=\mathcal{N}(x_i|0,\alpha_j^{-1})$ for
$j=i,i+1,i-1$. We see that the prior for $x_i$ is proportional to
a product of three Gaussian distributions, with the coefficient
$x_i$ associated with one of the three hyperparameters
$\{\alpha_i,\alpha_{i+1},\alpha_{i-1}\}$ for each distribution.
When $\beta=0$, the prior distribution (\ref{eq2}) reduces to the
prior for the conventional sparse Bayesian learning. When
$\beta>0$, the sparsity of $x_i$ not only depends on the
hyperparameter $\alpha_i$, but also on the neighboring
hyperparameters $\{\alpha_{i+1},\alpha_{i-1}\}$. Hence it can be
expected that the sparsity patterns of neighboring coefficients
are related to each other. Also, such a prior does not require the
knowledge of the block-sparse structure of the sparse signal. It
naturally has the tendency to suppress isolated non-zero
coefficients and encourage structured-sparse solutions.


Following the conventional sparse Bayesian learning framework, we
use Gamma distributions as hyperpriors over the hyperparameters
$\{\alpha_i\}$, i.e.
\begin{align}
p(\boldsymbol{\alpha})=\prod_{i=1}^n\text{Gamma}(\alpha_i|a,b)=\prod_{i=1}^n\Gamma(a)^{-1}b^a\alpha^{a}e^{-b\alpha}
\label{alpha-prior}
\end{align}
where $\Gamma(a)=\int_{0}^{\infty}t^{a-1}e^{-t}dt$ is the Gamma
function. The choice of the Gamma hyperprior results in a learning
process which tends to switch off most of the coefficients that
are deemed to be irrelevant, and only keep very few relevant
coefficients to explain the data. This mechanism is also called as
``automatic relevance determination''. In the conventional sparse
Bayesian framework, to make the Gamma prior non-informative, very
small values, e.g. $10^{-4}$, are assigned to the two parameters
$a$ and $b$. Nevertheless, in this paper, we use a more favorable
prior which sets a larger $a$ (say, $a=1$) in order to achieve the
desired ``pruning'' effect for our proposed hierarchical Bayesian
model. Clearly, the Gamma prior with a larger $a$ encourages large
values of the hyperparameters, and therefore promotes the
sparseness of the solution since the larger the hyperparameter,
the smaller the variance of the corresponding coefficient.








\section{Proposed Bayesian Inference Algorithm} \label{sec:inference}
We now proceed to develop a sparse Bayesian learning method for
block-sparse signal recovery. For ease of exposition, we assume
that the noise variance $\sigma^2$ is known \emph{a priori}.
Extension of the Bayesian inference to the case of unknown noise
variance will be discussed in the next section. Based on the above
hierarchical model, the posterior distribution of $\boldsymbol{x}$
can be computed as
\begin{align}
p(\boldsymbol{x}|\boldsymbol{\alpha},\boldsymbol{y})\propto&
p(\boldsymbol{x}|\boldsymbol{\alpha})p(\boldsymbol{y}|\boldsymbol{x})
\label{x-posterior}
\end{align}
where $\boldsymbol{\alpha}\triangleq\{\alpha_i\}$,
$p(\boldsymbol{x}|\boldsymbol{\alpha})$ is given by (\ref{eq3}),
and
\begin{align}
 p(\boldsymbol{y}|\boldsymbol{x})=&
\frac{1}{(\sqrt{2\pi\sigma^2})^{m}}\exp\bigg(-\frac{\|\boldsymbol{y}-\boldsymbol{A}\boldsymbol{x}\|_2^2}{2\sigma^2}\bigg)
\end{align}
It can be readily verified that the posterior
$p(\boldsymbol{x}|\boldsymbol{\alpha},\boldsymbol{y})$ follows a
Gaussian distribution with its mean and covariance given
respectively by
\begin{align}
\boldsymbol{\mu}=&\sigma^{-2}\boldsymbol{\Phi}\boldsymbol{A}^T\boldsymbol{y}
\nonumber\\
\boldsymbol{\Phi}=&(\sigma^{-2}\boldsymbol{A}^T\boldsymbol{A}+\boldsymbol{D})^{-1}
\label{eq4}
\end{align}
where $\boldsymbol{D}$ is a diagonal matrix with its $i$th
diagonal element equal to
$(\alpha_i+\beta\alpha_{i+1}+\beta\alpha_{i-1})$, i.e.
\begin{align}
\boldsymbol{D}\triangleq\text{diag}(\alpha_1+\beta\alpha_{2}+\beta\alpha_{0},\ldots,
\alpha_{n}+\beta\alpha_{n-1}+\beta\alpha_{n+1})
\label{D-definition}
\end{align}
Given a set of estimated hyperparameters $\{\alpha_i\}$, the
maximum a posterior (MAP) estimate of $\boldsymbol{x}$ is the mean
of its posterior distribution, i.e.
\begin{align}
\boldsymbol{\hat{x}}_{\text{MAP}}=\boldsymbol{\mu}=
(\boldsymbol{A}^T\boldsymbol{A}+\sigma^{2}\boldsymbol{D})^{-1}
\boldsymbol{A}^T\boldsymbol{y} \label{posterior-mean}
\end{align}

Our problem therefore reduces to estimating the set of
hyperparameters $\{\alpha_i\}$. With hyperpriors placed over
$\alpha_{i}$, learning the hyperparameters becomes a search for
their posterior mode, i.e. maximization of the posterior
probability $p(\boldsymbol{\alpha}|\boldsymbol{y})$. A strategy to
maximize the posterior probability is to exploit the
expectation-maximization (EM) formulation which treats the signal
$\boldsymbol{x}$ as the hidden variables and maximizes the
expected value of the complete log-posterior of
$\boldsymbol{\alpha}$, i.e.
$E_{\boldsymbol{x}|\boldsymbol{y},\boldsymbol{\alpha}}[\log
p(\boldsymbol{\alpha}|\boldsymbol{x})]$, where the operator
$E_{\boldsymbol{x}|\boldsymbol{y},\boldsymbol{\alpha}}[\cdot]$
denotes the expectation with respect to the distribution
$p(\boldsymbol{x}|\boldsymbol{y},\boldsymbol{\alpha})$.
Specifically, the EM algorithm produces a sequence of estimates
$\boldsymbol{\alpha}^{(t)}$, $t=1,2,3,\ldots$, by applying two
alternating steps, namely, the E-step and the M-step
\cite{FangJi08}.

\textbf{E-Step:} Given the current estimates of the
hyperparameters $\boldsymbol{\alpha}^{(t)}$ and the observed data
$\boldsymbol{y}$, the E-step requires computing the expected value
(with respect to the missing variables $\boldsymbol{x}$) of the
complete log-posterior of $\boldsymbol{\alpha}$, which is also
referred to as the Q-function; we have
\begin{align}
Q(\boldsymbol{\alpha}|\boldsymbol{\alpha}^{(t)})=&
E_{\boldsymbol{x}|\boldsymbol{y},\boldsymbol{\alpha}^{(t)}}[\log
p(\boldsymbol{\alpha}|\boldsymbol{x})] \nonumber\\
=&\int
p(\boldsymbol{x}|\boldsymbol{y},\boldsymbol{\alpha}^{(t)})\log
p(\boldsymbol{\alpha}|\boldsymbol{x}) d\boldsymbol{x} \nonumber\\
=&\int
p(\boldsymbol{x}|\boldsymbol{y},\boldsymbol{\alpha}^{(t)})\log
[p(\boldsymbol{\alpha})p(\boldsymbol{x}|\boldsymbol{\alpha})]d\boldsymbol{x}+
c \label{Q-function-org}
\end{align}
where $c$ is a constant independent of $\boldsymbol{\alpha}$.
Ignoring the term independent of $\boldsymbol{\alpha}$, and
recalling (\ref{eq3}), the Q-function can be re-expressed as
\begin{align}
Q(\boldsymbol{\alpha}|\boldsymbol{\alpha}^{(t)}) =& \log
p(\boldsymbol{\alpha})+\frac{1}{2}\sum_{i=1}^n\bigg(\log
(\alpha_i+\beta\alpha_{i+1}+\beta\alpha_{i-1}) \nonumber\\
&-(\alpha_i+\beta\alpha_{i+1}+\beta\alpha_{i-1})\int
p(\boldsymbol{x}|\boldsymbol{y},\boldsymbol{\alpha}^{(t)})x_i^2
d\boldsymbol{x}\bigg)
\end{align}
Since the posterior
$p(\boldsymbol{x}|\boldsymbol{y},\boldsymbol{\alpha}^{(t)})$ is a
multivariate Gaussian distribution with its mean and covariance
matrix given by (\ref{eq4}), we have
\begin{align}
\int
p(\boldsymbol{x}|\boldsymbol{y},\boldsymbol{\alpha}^{(t)})x_i^2
d\boldsymbol{x}=E_{\boldsymbol{x}|\boldsymbol{y},\boldsymbol{\alpha}^{(t)}}\left[x_i^2\right]=\hat{\mu}_i^2+\hat{\phi}_{i,i}
\end{align}
where $\hat{\mu}_i$ denotes the $i$th entry of
$\boldsymbol{\hat{\mu}}$, $\hat{\phi}_{i,i}$ denotes the $i$th
diagonal element of the covariance matrix
$\boldsymbol{\hat{\Phi}}$, $\boldsymbol{\hat{\mu}}$ and
$\boldsymbol{\hat{\Phi}}$ are computed according to (\ref{eq4}),
with $\boldsymbol{\alpha}$ replaced by the current estimate
$\boldsymbol{\alpha}^{(t)}$. With the specified prior
(\ref{alpha-prior}), the Q-function can eventually be written as
\begin{align}
&Q(\boldsymbol{\alpha}|\boldsymbol{\alpha}^{(t)})\nonumber\\
=&\sum_{i=1}^n\bigg(a\log\alpha_i-b\alpha_i+\frac{1}{2}\log
(\alpha_i+\beta\alpha_{i+1}+\beta\alpha_{i-1}) \nonumber\\
&-\frac{1}{2}(\alpha_i+\beta\alpha_{i+1}+\beta\alpha_{i-1})(\hat{\mu}_i^2+\hat{\phi}_{i,i})\bigg)
\label{Q-function}
\end{align}


\textbf{M-Step:} In the M-step of the EM algorithm, a new estimate
of $\boldsymbol{\alpha}$ is obtained by maximizing the Q-function,
i.e.
\begin{align}
\boldsymbol{\alpha}^{(t+1)} =
\arg\max_{\boldsymbol{\alpha}}Q(\boldsymbol{\alpha}|\boldsymbol{\alpha}^{(t)})
\label{M-opt}
\end{align}
For the conventional sparse Bayesian learning, maximization of the
Q-function can be decoupled into a number of separate
optimizations in which each hyperparameter $\alpha_i$ is updated
independently. This, however, is not the case for the problem
being considered here. We see that the hyperparameters in the
Q-function (\ref{Q-function}) are entangled with each other due to
the logarithm term
$\log(\alpha_i+\beta\alpha_{i+1}+\beta\alpha_{i-1})$. In this
case, an analytical solution to the optimization (\ref{M-opt}) is
difficult to obtain. Gradient descend methods can certainly be
used to search for the optimal solution. Nevertheless, such a
gradient-based search method, albeit effective, does not provide
any insight into the learning process. Also, gradient-based
methods involve higher computational complexity as compared with
an analytical update rule. To overcome the drawbacks of
gradient-based methods, we consider an alternative strategy which
aims at finding a simple, analytical sub-optimal solution of
(\ref{M-opt}). Such an analytical sub-optimal solution can be
obtained by examining the optimality condition of (\ref{M-opt}).
Suppose $\boldsymbol{\alpha}^{\ast}$ is the optimal solution of
(\ref{M-opt}), then the first derivative of the Q-function with
respect to $\boldsymbol{\alpha}$ equals to zero at the optimal
point, i.e.
\begin{align}
\frac{\partial
Q(\boldsymbol{\alpha}|\boldsymbol{\alpha}^{(t)})}{\partial\boldsymbol{\alpha}}
\bigg
|_{\boldsymbol{\alpha}=\boldsymbol{\alpha}{\ast}}=\boldsymbol{0}
\end{align}
To examine this optimality condition more thoroughly, we compute
the first derivative of the Q-function with respect to each
individual hyperparameter:
\begin{align}
\frac{\partial
Q(\boldsymbol{\alpha}|\boldsymbol{\alpha}^{(t)})}{\partial\alpha_i}
=\frac{a}{\alpha_i}-b-\frac{1}{2}\omega_i
+\frac{1}{2}(\nu_i+&\beta\nu_{i+1}+\beta\nu_{i-1})\nonumber\\
& \quad \forall i=1,\ldots,n
\end{align}
where $\nu_0=0$, $\nu_{n+1}=0$, and for $i=1,\ldots,n$, we have
\begin{align}
\omega_i\triangleq&(\hat{\mu}_i^2+\hat{\phi}_{i,i})+\beta(\hat{\mu}_{i+1}^2+\hat{\phi}_{i+1,i+1})
+\beta(\hat{\mu}_{i-1}^2+\hat{\phi}_{i-1,i-1}) \label{omega} \\
\nu_i\triangleq&\frac{1}{\alpha_i+\beta\alpha_{i+1}+\beta\alpha_{i-1}}
\end{align}
Note that for notational convenience, we allow the subscript
indices of the notations $\hat{\mu}_i$ and $\hat{\phi}_{i,i}$ in
(\ref{omega}) equal to $0$ and $n+1$. Although these notations
$\{\hat{\mu}_0, \hat{\phi}_{0,0},\hat{\mu}_{n+1},
\hat{\phi}_{n+1,n+1}\}$ do not have any meaning, they can be used
to simplify our expression. Clearly, they should all be set equal
to zero, i.e.
$\hat{\mu}_0=\hat{\mu}_{n+1}=\hat{\phi}_{0,0}=\hat{\phi}_{n+1,n+1}=0$.
Recalling the optimality condition, we therefore have
\begin{align}
\frac{a}{\alpha_i^{\ast}}+
\frac{1}{2}(\nu_i^{\ast}+\beta\nu_{i+1}^{\ast}+\beta\nu_{i-1}^{\ast})
=b+\frac{1}{2}\omega_i \quad \forall i=1,\ldots,n\label{eq5}
\end{align}
where $\nu_0^{\ast}=0$, $\nu_{n+1}^{\ast}=0$, and
\begin{align}
\nu_i^{\ast}\triangleq&\frac{1}{\alpha_i^{\ast}+\beta\alpha_{i+1}^{\ast}+\beta\alpha_{i-1}^{\ast}}\qquad
\forall i=1,\ldots,n \nonumber
\end{align}
Since all hyperparameters $\{\alpha_i\}$ and $\beta$ are
non-negative, we have
\begin{align}
\frac{1}{\alpha_i^{\ast}}>&\nu_i^{\ast}>0 \qquad \forall
i=1,\ldots,n \nonumber\\
\frac{1}{\beta\alpha_{i+1}^{\ast}}>&\nu_i^{\ast}>0 \qquad
\forall i=1,\ldots,n-1 \nonumber\\
\frac{1}{\beta\alpha_{i-1}^{\ast}}>&\nu_i^{\ast}>0 \qquad \forall
i=2,\ldots,n \nonumber
\end{align}
Hence the term on the left-hand side of (\ref{eq5}) is lower and
upper bounded respectively by
\begin{align}
\frac{a+c_0}{\alpha_i^{\ast}}\geq\frac{a}{\alpha_i^{\ast}}+
\frac{1}{2}(\nu_i^{\ast}+\beta\nu_{i+1}^{\ast}+\beta\nu_{i-1}^{\ast})>\frac{a}{\alpha_i^{\ast}}
\label{eq6}
\end{align}
where $c_0=1.5$ for $i=2,\ldots,n-1$, and $c_0=1$ for $i=\{1,n\}$.
Combining (\ref{eq5})--(\ref{eq6}), we arrive at
\begin{align}
\alpha_i^{\ast}\in \left[\frac{a}{0.5\omega_i+b},
\frac{a+c_0}{0.5\omega_i+b}\right]\quad \forall i=1,\ldots,n
\label{eq12}
\end{align}
With $a=1$, and $b=10^{-4}$, a sub-optimal solution to
(\ref{M-opt}) can be obtained as
\begin{align}
\hat{\alpha}_i=\frac{\kappa}{0.5\omega_i+10^{-4}} \quad \forall
i=1,\ldots,n \label{hp-update}
\end{align}
for some $\kappa$ within the range $1+c_0\geq\kappa\geq 1$. We see
that the solution (\ref{hp-update}) provides a simple rule for the
hyperparameter update. Also, notice that the update rule
(\ref{hp-update}) resembles that of the conventional sparse
Bayesian learning work \cite{Tipping01,JiXue08} except that the
parameter $\omega_i$ is equal to $\hat{\mu}_i^2+\hat{\phi}_{i,i}$
for the conventional sparse Bayesian learning method, while for
our case, $\omega_i$ is a weighted summation of
$\hat{\mu}_j^2+\hat{\phi}_{j,j}$ for $j=i-1,i,i+1$.


For clarity, we now summarize the EM algorithm as follows.
\begin{enumerate}
\item At iteration $t$ ($t=0,1,\ldots$): Given a set of hyperparameters
$\boldsymbol{\alpha}^{(t)}=\{\alpha_i^{(t)}\}$, compute the mean
$\boldsymbol{\hat{\mu}}$ and covariance matrix
$\boldsymbol{\hat{\Phi}}$ of the posterior distribution
$p(\boldsymbol{x}|\boldsymbol{\alpha}^{(t)},\boldsymbol{y})$
according to (\ref{eq4}), and compute the MAP estimate
$\boldsymbol{\hat{x}}^{(t)}$ according to (\ref{posterior-mean}).
\item Update the hyperparameters $\boldsymbol{\alpha}^{(t+1)}$ according to
(\ref{hp-update}), where $\omega_i$ is given by (\ref{omega}).
\item Continue the above iteration until $\|\boldsymbol{\hat{x}}^{(t+1)}-\boldsymbol{\hat{x}}^{(t)}\|_2\leq\epsilon$,
where $\epsilon$ is a prescribed tolerance value.
\end{enumerate}


\emph{Remarks:}  Although the above algorithm employs a
sub-optimal solution (\ref{hp-update}) to update the
hyperparameters in the M-step, numerical results show that the
sub-optimal update rule is quite effective and presents similar
recovery performance as using a gradient-based search method. This
is because the sub-optimal solution (\ref{hp-update}) provides a
reasonable estimate of the optimal solution when the parameter $a$
is set away from zero, say, $a=1$. Numerical results also suggest
that the proposed algorithm is insensitive to the choice of the
parameter $\kappa$ in (\ref{hp-update}) as long as $\kappa$ is
within the range $[a, a+c_0]$ for a properly chosen $a$. We simply
set $\kappa=a$ in our following simulations.



The update rule (\ref{hp-update}) not only admits a simple
analytical form which is computationally efficient, it also
provides an insight into the EM algorithm. The Bayesian Occam's
razor which contributes to the success of the conventional sparse
Bayesian learning method also works here to automatically select
an appropriate simple model. To see this, note that in the E-step,
when computing the posterior mean and covariance matrix, a large
hyperparameter $\alpha_i$ tends to suppress the values of the
corresponding components $\{\mu_j,\phi_j\}$ for $j=i-1,i,i+1$
(c.f. (\ref{eq4})). As a result, the value of $\omega_i$ becomes
small, which in turn leads to a larger hyperparameter $\alpha_i$
(c.f. (\ref{hp-update})). This negative feedback mechanism keeps
decreasing most of the entries in $\boldsymbol{\hat{x}}$ until
they reach machine precision and become zeros, while leaving only
a few prominent nonzero entries survived to explain the data.
Meanwhile, we see that each hyperparameter $\alpha_i$ not only
controls the sparseness of its own corresponding coefficient
$x_i$, but also has an impact on the sparseness of the neighboring
coefficients $\{x_{i+1},x_{i-1}\}$. Therefore the proposed EM
algorithm has the tendency to suppress isolated non-zero
coefficients and encourage structured-sparse solutions.

\section{Bayesian Inference: Unknown Noise Variance} \label{sec:inference-un}
In the previous section, for simplicity of exposition, we assume
that the noise variance $\sigma^2$ is known \emph{a priori}. This
assumption, however, may not hold valid in practice. In this
section, we discuss how to extend our previously developed
Bayesian inference method to the scenario where the noise variance
$\sigma^2$ is unknown.

For notational convenience, define
\begin{align}
\gamma\triangleq\sigma^{-2} \nonumber
\end{align}
Following the conventional sparse Bayesian learning framework
\cite{Tipping01}, we place a Gamma hyperprior over $\gamma$:
\begin{align}
p(\gamma)=\text{Gamma}(\gamma|c,d)=\Gamma(c)^{-1}d^c\gamma^{c}e^{-d\gamma}
\label{gamma-prior}
\end{align}
where the parameters $c$ and $d$ are set to small values, e.g.
$c=d=10^{-4}$. As we already derived in the previous section,
given the hyperparameters $\boldsymbol{\alpha}$ and the noise
variance $\sigma^2$, the posterior
$p(\boldsymbol{x}|\boldsymbol{\alpha},\gamma,\boldsymbol{y})$
follows a Gaussian distribution with its mean and covariance
matrix given by (\ref{eq4}). The MAP estimate of $\boldsymbol{x}$
is equivalent to the posterior mean. Our problem therefore becomes
jointly estimating the hyperparameters $\boldsymbol{\alpha}$ and
the noise variance $\sigma^2$ (or equivalently $\gamma$). Again,
the EM algorithm can be used to learn these parameters via
maximizing their posterior probability
$p(\boldsymbol{\alpha},\gamma|\boldsymbol{y})$. The alternating EM
steps are briefly discussed below.

\textbf{E-Step}: In the E-step, given the current estimates of the
parameters $\{\boldsymbol{\alpha}^{(t)},\gamma^{(t)}\}$ and the
observed data $\boldsymbol{y}$, we compute the expected value
(with respect to the missing variables $\boldsymbol{x}$) of the
complete log-posterior of $\{\boldsymbol{\alpha},\gamma\}$, that
is,
$E_{\boldsymbol{x}|\boldsymbol{y},\boldsymbol{\alpha}^{(t)},\gamma^{(t)}}[\log
p(\boldsymbol{\alpha},\gamma|\boldsymbol{x},\boldsymbol{y})]$,
where the operator
$E_{\boldsymbol{x}|\boldsymbol{y},\boldsymbol{\alpha}^{(t)},\gamma^{(t)}}[\cdot]$
denotes the expectation with respect to the distribution
$p(\boldsymbol{x}|\boldsymbol{y},\boldsymbol{\alpha}^{(t)},\gamma^{(t)})$.
Since
\begin{align}
p(\boldsymbol{\alpha},\gamma|\boldsymbol{x},\boldsymbol{y})\propto
p(\boldsymbol{\alpha})p(\boldsymbol{x}|\boldsymbol{\alpha})p(\gamma)p(\boldsymbol{y}|\boldsymbol{x},\gamma)
\end{align}
the Q-function can be expressed as a summation of two terms
\begin{align}
Q(\boldsymbol{\alpha},\gamma|\boldsymbol{\alpha}^{(t)},\gamma^{(t)})=&
E_{\boldsymbol{x}|\boldsymbol{y},\boldsymbol{\alpha}^{(t)},\gamma^{(t)}}[\log
p(\boldsymbol{\alpha})p(\boldsymbol{x}|\boldsymbol{\alpha})]
\nonumber\\
&+
E_{\boldsymbol{x}|\boldsymbol{y},\boldsymbol{\alpha}^{(t)},\gamma^{(t)}}[\log
p(\gamma)p(\boldsymbol{y}|\boldsymbol{x},\gamma)] \label{eq7}
\end{align}
where the first term has exactly the same form as the Q-function
(\ref{Q-function-org}) obtained in the previous section, except
with the known noise variance $\sigma^2$ replaced by the current
estimate $(\sigma^{(t)})^2=1/\gamma^{(t)}$, and the second term is
a function of the variable $\gamma$.

\textbf{M-Step}: We observe that in the Q-function (\ref{eq7}),
the parameters $\boldsymbol{\alpha}$ and $\gamma$ to be learned
are separated from each other. This allows the estimation of
$\boldsymbol{\alpha}$ and $\gamma$ to be decoupled into the
following two independent problems:
\begin{align}
\boldsymbol{\alpha}^{(t+1)} =&
\arg\max_{\boldsymbol{\alpha}}E_{\boldsymbol{x}|\boldsymbol{y},\boldsymbol{\alpha}^{(t)},\gamma^{(t)}}[\log
p(\boldsymbol{\alpha})p(\boldsymbol{x}|\boldsymbol{\alpha})]
\label{opt-2}
\\
\gamma^{(t+1)} =& \arg\max_{\gamma}
E_{\boldsymbol{x}|\boldsymbol{y},\boldsymbol{\alpha}^{(t)},\gamma^{(t)}}[\log
p(\gamma)p(\boldsymbol{y}|\boldsymbol{x},\gamma)] \label{opt-3}
\end{align}
The first optimization problem (\ref{opt-2}) has been thoroughly
studied in the previous section, where we provided a simple
analytical form (\ref{hp-update}) for the hyperparameter update.
We now discuss the estimation of the parameter $\gamma$. Recalling
(\ref{gamma-prior}), we have
\begin{align}
&E_{\boldsymbol{x}|\boldsymbol{y},\boldsymbol{\alpha}^{(t)},\gamma^{(t)}}[\log
p(\gamma)p(\boldsymbol{y}|\boldsymbol{x},\gamma)] \nonumber\\
=&\frac{m}{2}\log\gamma-\frac{\gamma}{2}E_{\boldsymbol{x}|\boldsymbol{y},\boldsymbol{\alpha}^{(t)},\gamma^{(t)}}
\left[\|\boldsymbol{y}-\boldsymbol{A}\boldsymbol{x}\|_2^2\right]+c\log\gamma-d\gamma
\label{eq13}
\end{align}
Computing the first derivative of (\ref{eq13}) with respect to
$\gamma$ and setting it equal to zero, we get
\begin{align}
\frac{1}{\gamma}=\frac{\chi+2d}{m+2c} \label{eq11}
\end{align}
where
\begin{align}
\chi\triangleq
E_{\boldsymbol{x}|\boldsymbol{y},\boldsymbol{\alpha}^{(t)},\gamma^{(t)}}
\left[\|\boldsymbol{y}-\boldsymbol{A}\boldsymbol{x}\|_2^2\right]
\nonumber
\end{align}
Note that the posterior
$p(\boldsymbol{x}|\boldsymbol{y},\boldsymbol{\alpha}^{(t)},\gamma^{(t)})$
follows a Gaussian distribution with mean $\boldsymbol{\hat{\mu}}$
and covariance matrix $\boldsymbol{\hat{\Phi}}$, where
$\boldsymbol{\hat{\mu}}$ and $\boldsymbol{\hat{\Phi}}$ are
computed via (\ref{eq4}) with $\gamma$ (i.e. $\sigma^2$) and
$\boldsymbol{\alpha}$ replaced by the current estimates
$\{\gamma^{(t)},\boldsymbol{\alpha}^{(t)}\}$. Hence $\chi$ can be
computed as
\begin{align}
\chi=&\boldsymbol{y}^T\boldsymbol{y}-2
E[\boldsymbol{x}^T\boldsymbol{A}^T\boldsymbol{y}]+E[\boldsymbol{x}^T\boldsymbol{A}^T
\boldsymbol{A}\boldsymbol{x}] \nonumber\\
=&\boldsymbol{y}^T\boldsymbol{y}-2\boldsymbol{\hat{\mu}}^T\boldsymbol{A}^T\boldsymbol{y}+
\boldsymbol{\hat{\mu}}^T\boldsymbol{A}^T
\boldsymbol{A}\boldsymbol{\hat{\mu}}+\text{tr}\left(\boldsymbol{\hat{\Phi}}\boldsymbol{A}^T\boldsymbol{A}\right)
\nonumber\\
\stackrel{(a)}{=}&
\|\boldsymbol{y}-\boldsymbol{A}\boldsymbol{\hat{\mu}}\|_2^2+(\gamma^{(t)})^{-1}\sum_{i=1}^n\rho_i
\label{eq10}
\end{align}
where the last equality $(a)$ follows from
\begin{align}
\text{tr}\left(\boldsymbol{\hat{\Phi}}\boldsymbol{A}^T\boldsymbol{A}\right)=&
\text{tr}\left(\boldsymbol{\hat{\Phi}}\boldsymbol{A}^T\boldsymbol{A}+(\gamma^{(t)})^{-1}
\boldsymbol{\hat{\Phi}}\boldsymbol{\hat{D}}-(\gamma^{(t)})^{-1}
\boldsymbol{\hat{\Phi}}\boldsymbol{\hat{D}}\right) \nonumber\\
=&(\gamma^{(t)})^{-1}\text{tr}\left(\boldsymbol{\hat{\Phi}}(\gamma^{(t)}\boldsymbol{A}^T\boldsymbol{A}+\boldsymbol{\hat{D}})
- \boldsymbol{\hat{\Phi}}\boldsymbol{\hat{D}}\right) \nonumber\\
=&(\gamma^{(t)})^{-1}\text{tr}\left(\boldsymbol{I}-\boldsymbol{\hat{\Phi}}\boldsymbol{\hat{D}}\right)
\nonumber\\
=&(\gamma^{(t)})^{-1}\sum_{i=1}^n\rho_i
\end{align}
in which $\boldsymbol{\hat{D}}$ is given by (\ref{D-definition})
with $\boldsymbol{\alpha}$ replaced by the current estimate
$\boldsymbol{\alpha}^{(t)}$, and
\begin{align}
\rho_i\triangleq
1-\hat{\phi}_{i,i}(\alpha_{i}^{(t)}+\beta\alpha_{i-1}^{(t)}+\beta\alpha_{i+1}^{(t)})
\qquad \forall i \label{rho}
\end{align}
Note that $\alpha_{0}^{(t)}$ and $\alpha_{n+1}^{(t)}$ are set to
zero when computing $\rho_1$ and $\rho_n$. Substituting
(\ref{eq10}) back into (\ref{eq11}), a new estimate of $\gamma$,
i.e. the optimal solution to (\ref{opt-3}), is given by
\begin{align}
\frac{1}{\gamma^{(t+1)}}=\frac{\|\boldsymbol{y}-\boldsymbol{A}\boldsymbol{\hat{\mu}}\|_2^2
+(\gamma^{(t)})^{-1}\sum_{i}\rho_i+2d}{m+2c} \label{gamma-update}
\end{align}
The above update formula has a similar form as that for the
conventional sparse Bayesian learning (c.f. \cite[Equation
(50)]{Tipping01}). The only difference lies in that $\{\rho_i\}$
are computed differently: for the conventional sparse Bayesian
learning method, $\rho_i$ is computed as
$\rho_i=1-\hat{\phi}_{i,i}\alpha_{i}^{(t)}$, while $\rho_i$ is
given by (\ref{rho}) for our algorithm.

The sparse Bayesian learning algorithm with unknown noise variance
is now summarized as follows.
\begin{enumerate}
\item At iteration $t$ ($t=0,1,\ldots$): given the current estimates of
$\boldsymbol{\alpha}^{(t)}$ and $\gamma^{(t)}$, compute the mean
$\boldsymbol{\hat{\mu}}$ and the covariance matrix
$\boldsymbol{\hat{\Phi}}$ of the posterior distribution
$p(\boldsymbol{x}|\boldsymbol{\alpha}^{(t)},\gamma^{(t)},\boldsymbol{y})$
via (\ref{eq4}), and calculate the MAP estimate
$\boldsymbol{\hat{x}}^{(t)}$ according to (\ref{posterior-mean}).
\item Compute a new estimate of $\boldsymbol{\alpha}$, denoted as $\boldsymbol{\alpha}^{(t+1)}$, according to
(\ref{hp-update}), where $\omega_i$ is given by (\ref{omega});
update $\gamma$ via (\ref{gamma-update}), which yields a new
estimate of $\gamma$, denoted as $\gamma^{(t+1)}$.
\item Continue the above iteration until $\|\boldsymbol{\hat{x}}^{(t+1)}-\boldsymbol{\hat{x}}^{(t)}\|_2\leq\epsilon$,
where $\epsilon$ is a prescribed tolerance value.
\end{enumerate}

\section{Discussions} \label{sec:discussions}
\subsection{Related Work}
Sparse Bayesian learning is a powerful approach for
regression, classification, and sparse representation. It was
firstly introduced by Tipping in his pioneering work
\cite{Tipping01}, where the regression and classification problem
was addressed and a sparse Bayesian learning approach was
developed to automatically remove irrelevant basis vectors and
retain only a few `relevant' vectors for prediction. Such an
automatic relevance determination mechanism and the resulting
sparse solution not only effectively avoid the overfitting
problem, but also render superior regression and classification
accuracy. Later on in \cite{Wipf06,JiXue08}, sparse Bayesian
learning was introduced to solve the sparse recovery problem. In a
series of experiments, sparse Bayesian learning demonstrated
superior stability for sparse signal recovery, and presents
uniform superiority over other methods.

In \cite{WipfRao07}, sparse Bayesian learning was generalized to
solve the simultaneous (block) sparse recovery problem, in which a
group of coefficients sharing the same sparsity pattern are
assigned a multivariate Gaussian prior parameterized by a common
hyperparameter that controls the sparsity of this group of
coefficients. Specifically, we have
\begin{align}
p(\boldsymbol{x}_i|\alpha_i)=\mathcal{N}(0,\alpha_i^{-1}\boldsymbol{I})
\end{align}
where $\boldsymbol{x}_i$ denotes the group of coefficients that
share a same sparsity pattern, $\alpha_i$ is the hyperparameter
controlling the sparsity of $\boldsymbol{x}_i$. In
\cite{ZhangRao11}, the above model was further improved to
accommodate temporally correlated sources
\begin{align}
p(\boldsymbol{x}_i|\alpha_i)=\mathcal{N}(0,\alpha_i^{-1}\boldsymbol{B}_i)
\end{align}
in which $\boldsymbol{B}_i$ is a positive definite matrix that
captures the correlation structure of $\boldsymbol{x}_i$. We see
that, in both models \cite{WipfRao07,ZhangRao11}, each coefficient
is associated with only one sparseness-controlling hyperparameter.
This explicit assignment of each coefficient to a certain
hyperparameter requires to know the exact block sparsity pattern
\emph{a priori}. In contrast, for our hierarchical Bayesian model,
each coefficient is associated with multiple hyperparameters, and
the hyperparameters are somehow related to each other through
their commonly connected coefficients. Such a coupled hierarchical
model has the potential to encourage block-sparse patterns, while
without imposing any stringent or pre-specified constraints on the
structure of the recovered signals. This property enables the
proposed algorithm to learn the block-sparse structure in an
automatic manner.

Recently, Zhang and Rao extended the block sparse Bayesian
learning framework to address the sparse signal recovery problem
when the block structure is unknown \cite{ZhangRao13}. In their
work \cite{ZhangRao13}, the signal $\boldsymbol{x}$ is partitioned
into a number of overlapping blocks $\{\boldsymbol{x}_i\}$ with
identical block sizes, and each block $\boldsymbol{x}_i$ is
assigned a Gaussian prior
$p(\boldsymbol{x}_i|\alpha_i)=\mathcal{N}(0,\alpha_i^{-1}\boldsymbol{B}_i)$.
To address the overlapping issue, the original data model is
converted into an expanded model which removes the overlapping
structure by adding redundant columns to the original measurement
matrix $\boldsymbol{A}$ and stacking all blocks
$\{\boldsymbol{x}_i\}$ to form an augmented vector. In doing this
way, the prior for the new augmented vector has a block diagonal
form similar as that for the conventional block sparse Bayesian
learning. Thus conventional block sparse Bayesian learning
algorithms such as \cite{ZhangRao11} can be applied to the
expanded model. This overlapping structure provides flexibility in
defining a block-sparse pattern. Hence it works well even when the
block structure is unknown. A critical difference between our work
and \cite{ZhangRao11} is that for our method, a prior is directly
placed on the signal $\boldsymbol{x}$, while for the method
proposed in \cite{ZhangRao11}, a rigorous formulation of the prior
for $\boldsymbol{x}$ is not available, instead, a prior is
assigned to the augmented new signal which is constructed by
stacking a number of overlapping blocks $\{\boldsymbol{x}_i\}$.

\subsection{A Proposed Iterative Reweighted Algorithm}
Sparse Bayesian learning algorithms have a close connection with
the reweighted $\ell_1$ or $\ell_2$ methods. In fact, a dual-form
analysis \cite{WipfNagaranjan10} reveals that sparse Bayesian
learning can be considered as a non-separable reweighted strategy
solving a non-separable penalty function. Inspired by this
insight, we here propose a reweighted $\ell_1$ method for the
recovery of block-sparse signals when the block structure of the
sparse signal is unknown.

Conventional reweighted $\ell_1$ methods iteratively minimize the
following weighted $\ell_1$ function (for simplicity, we consider
the noise-free case):
\begin{align}
\min_{\boldsymbol{x}}\quad& \sum_{i=1}^n
w_i^{(t)}|x_i|\nonumber\\
\text{s.t.}\quad&
\phantom{0}\boldsymbol{Ax}=\boldsymbol{y}\label{opt-2}
\end{align}
where the weighting parameters are given by
$w_i^{(t)}=1/(|x_i^{(t-1)}|+\epsilon),\forall i$, and $\epsilon$
is a pre-specified positive parameter. In a series of experiments
\cite{CandesWakin08}, the above iterative reweighted algorithm
outperforms the conventional $\ell_1$-minimization method by a
considerable margin. The fascinating idea of the iterative
reweighted algorithm is that the weights are updated based on the
previous estimate of the solution, with a large weight assigned to
the coefficient whose estimate is already small and vice versa. As
a result, the value of the coefficient which is assigned a large
weight tends to be smaller (until become negligible) in the next
estimate. This explains why iterative reweighted algorithms
usually yield sparser solutions than the conventional
$\ell_1$-minimization method.

As discussed in our previous section, the basic idea of our
proposed sparse Bayesian learning method is to establish a
coupling mechanism such that the sparseness of neighboring
coefficients are somehow related to each other. With this in mind,
we slightly modify the weight update rule of the reweighted
$\ell_1$ algorithm as follows
\begin{align}
w_i^{(t)}=\frac{1}{|x_i^{(t-1)}|+\beta|x_{i+1}^{(t-1)}|+\beta|x_{i-1}^{(t-1)}|+\epsilon}
\qquad \forall i
\end{align}
We see that unlike the conventional update rule, the weight
$w_i^{(t)}$ is not only a function of its corresponding
coefficient $x_i^{(t-1)}$, but also dependent on the neighboring
coefficients $\{x_{i+1}^{(t-1)},x_{i-1}^{(t-1)}\}$. In doing this
way, a coupling effect between the sparsity patterns of
neighboring coefficients is established. Hence the modified
reweighted $\ell_1$-minimization algorithm has the potential to
encourage block-sparse solutions. Experiments show that the
proposed modified reweighted $\ell_1$ method yields considerably
improved results over the conventional reweighted $\ell_1$ method
in recovering block-sparse signals. It also serves as a good
reference method for comparison with the proposed Bayesian sparse
learning approach.








\section{Simulation Results} \label{sec:simulation}
We now carry out experiments to illustrate the performance of our
proposed algorithm, also referred to as the pattern-coupled sparse
Bayesian learning (PC-SBL) algorithm, and its comparison with
other existing methods. The performance of the proposed algorithm
will be examined using both synthetic and real
data\footnote{Matlab codes for our algorithm are available at
http://www.junfang-uestc.net/}. The parameters $a$ and $b$ for our
proposed algorithm are set equal to $a=0.5$ and $b=10^{-4}$
throughout our experiments.


\subsection{Synthetic Data}
Let us first consider the synthetic data case. In our simulations,
we generate the block-sparse signal in a similar way to
\cite{ZhangRao13}. Suppose the $n$-dimensional sparse signal
contains $K$ nonzero coefficients which are partitioned into $L$
blocks with random sizes and random locations. Specifically, the
block sizes $\{B_l\}_{l=1}^L$ can be determined as follows: we
randomly generate $L$ positive random variables $\{r_l\}_{l=1}^L$
with their sum equal to one, then we can simply set $B_l=\lceil K
r_l\rceil$ for the first $L-1$ blocks and
$B_L=K-\sum_{l=1}^{L-1}B_l$ for the last block, where $\lceil
x\rceil$ denotes the ceiling operator that gives the smallest
integer no smaller than $x$. Similarly, we can partition the
$n$-dimensional vector into $L$ super-blocks using the same set of
values $\{r_l\}_{l=1}^L$, and place each of the $L$ nonzero blocks
into each super-block with a randomly generated starting position
(the starting position, however, is selected such that the nonzero
block will not go beyond the super-block). Also, in our
experiments, the nonzero coefficients of the sparse signal
$\boldsymbol{x}$ and the measurement matrix
$\boldsymbol{A}\in\mathbb{R}^{m\times n}$ are randomly generated
with each entry independently drawn from a normal distribution,
and then the sparse signal $\boldsymbol{x}$ and columns of
$\boldsymbol{A}$ are normalized to unit norm.



Two metrics are used to evaluate the recovery performance of
respective algorithms, namely, the normalized mean squared error
(NMSE) and the success rate. The NMSE is defined as
$\|\boldsymbol{x}-\boldsymbol{\hat{x}}\|_2^2/\|\boldsymbol{x}\|_2^2$,
where $\boldsymbol{\hat{x}}$ denotes the estimate of the true
signal $\boldsymbol{x}$. The success rate is computed as the ratio
of the number of successful trials to the total number of
independent runs. A trial is considered successful if the NMSE is
no greater than $10^{-4}$. In our simulations, the success rate is
used to measure the recovery performance for the noiseless case,
while the NMSE is employed to measure the recovery accuracy when
the measurements are corrupted by additive noise.

\begin{figure}[!t]
\centering
\includegraphics[width=9cm]{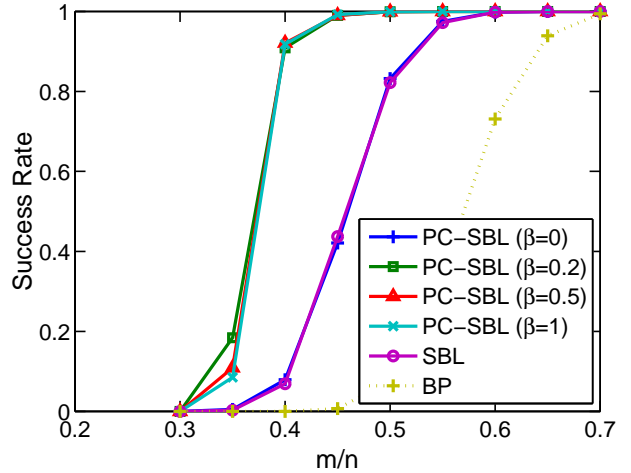}
\caption{Success rates of the proposed algorithm vs. the ratio
$m/n$ for different choices of $\beta$.} \label{fig1}
\end{figure}

We first examine the recovery performance of our proposed
algorithm (PC-SBL) under different choices of $\beta$. As
indicated earlier in our paper, $\beta$ ($0\leq\beta\leq 1$) is a
parameter quantifying the dependencies among neighboring
coefficients. Fig. \ref{fig1} depicts the success rates vs. the
ratio $m/n$ for different choices of $\beta$, where we set
$n=100$, $K=25$, and $L=4$. Results (in Fig. \ref{fig1} and the
following figures) are averaged over 1000 independent runs, with
the measurement matrix and the sparse signal randomly generated
for each run. The performance of the conventional sparse Bayesian
learning method (denoted as ``SBL'') \cite{Tipping01} and the
basis pursuit method (denoted as ``BP'')
\cite{ChenDonoho98,CandesTao05} is also included for our
comparison. We see that when $\beta=0$, our proposed algorithm
performs the same as the SBL. This is an expected result since in
the case of $\beta=0$, our proposed algorithm is simplified as the
SBL. Nevertheless, when $\beta>0$, our proposed algorithm achieves
a significant performance improvement (as compared with the SBL
and BP) through exploiting the underlying block-sparse structure,
even without knowing the exact locations and sizes of the non-zero
blocks. We also observe that our proposed algorithm is not very
sensitive to the choice of $\beta$ as long as $\beta>0$: it
achieves similar success rates for different positive values of
$\beta$. For simplicity, we set $\beta=1$ throughout our following
experiments.

\begin{figure*}[!t]
 \centering
\subfigure[Success rates vs. $m/n$, $n=100$, $K=25$, and
$L=4$]{\includegraphics[width=9cm]{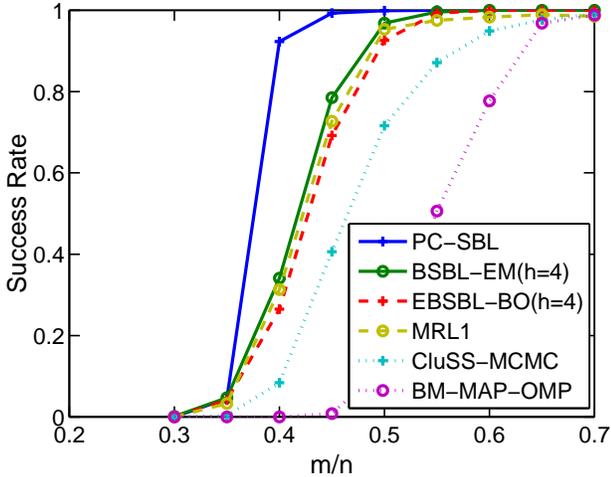}}
 \hfil
\subfigure[Success rates vs. the sparsity level $K$, $m=40$,
$n=100$, and
$L=3$]{\includegraphics[width=9cm]{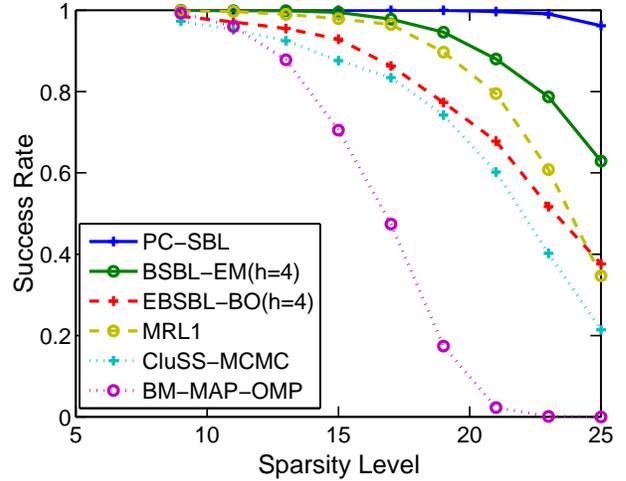}}
  \caption{Success rates of respective algorithms.}
   \label{fig2}
\end{figure*}

Next, we compare our proposed algorithm with some other recently
developed algorithms for block-sparse signal recovery, namely, the
expanded block sparse Bayesian learning method (EBSBL)
\cite{ZhangRao13}, the Boltzman machine-based greedy pursuit
algorithm (BM-MAP-OMP) \cite{PelegEldar12}, and the
cluster-structured MCMC algorithm (CluSS-MCMC) \cite{YuSun12}. The
modified iterative reweighted $\ell_1$ method (denoted as MRL1)
proposed in Section \ref{sec:discussions} is also examined in our
simulations. Note that all these algorithms were developed without
the knowledge of the block-sparse structure. The block sparse
Bayesian learning method (denoted as BSBL) developed in
\cite{ZhangRao13} is included as well. Although the BSBL algorithm
requires the knowledge of the block-sparse structure, it still
provides decent performance if the presumed block size, denoted by
$h$, is properly selected. Fig. \ref{fig2} plots the success rates
of respective algorithms as a function of the ratio $m/n$ and the
sparsity level $K$, respectively. Simulation results show that our
proposed algorithm achieves highest success rates among all
algorithms and outperforms other methods by a considerable margin.
We also noticed that the modified reweighted $\ell_1$ method
(MRL1), although not as good as the proposed PD-SBL, still
delivers acceptable performance which is comparable to the BSBL
and better than the BM-MAP-OMP and the CluSS-MCMC.


\begin{figure*}[!t]
 \centering
\subfigure[Normalized MSEs vs. $m/n$, $n=100$, $K=25$, and
$L=4$]{\includegraphics[width=9cm]{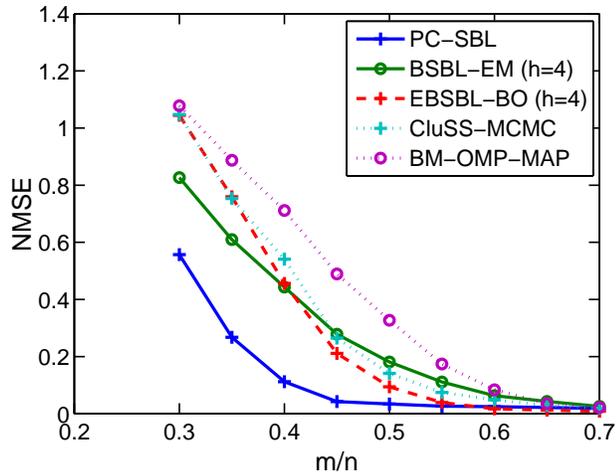}}
 \hfil
\subfigure[Normalized MSEs vs. the sparsity level $K$, $m=40$,
$n=100$, and $L=3$]{\includegraphics[width=9cm]{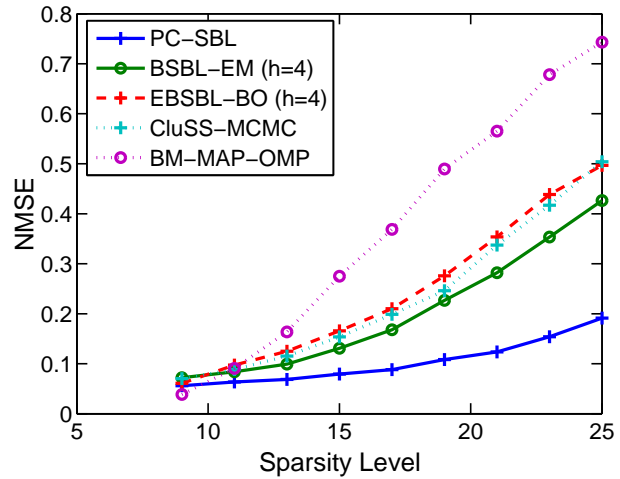}}
  \caption{Normalized MSEs of respective algorithms.}
   \label{fig3}
\end{figure*}

We now consider the noisy case where the measurements are
contaminated by additive noise. The observation noise is assumed
multivariate Gaussian with zero mean and covariance matrix
$\sigma^2\boldsymbol{I}$. Also, in our simulations, the noise
variance is assumed unknown (except for the BM-MAP-OMP). The NMSEs
of respective algorithms as a function of the ratio $m/n$ and the
sparsity level $K$ are plotted in Fig. \ref{fig3}, where the white
Gaussian noise is added such that the signal-to-noise ratio (SNR),
which is defined as $\text{SNR(dB)}\triangleq
20\log_{10}(\|\boldsymbol{Ax}\|_2/\|\boldsymbol{w}\|_2)$, is equal
to $15$dB for each iteration. We see that our proposed algorithm
yields a lower estimation error than other methods in the presence
of additive Gaussian noise.


\subsection{Real Data}
\begin{figure*}[!t]
 \centering
\begin{tabular}{cccccc}
\hspace*{-3ex}
\includegraphics[width=5cm]{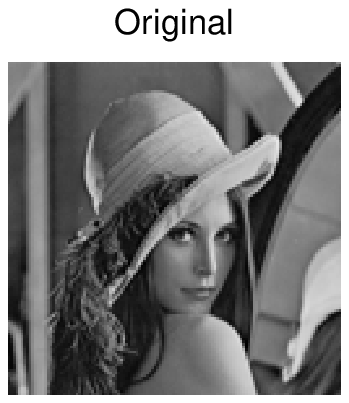}&
\hspace*{-10ex}
\includegraphics[width=5cm]{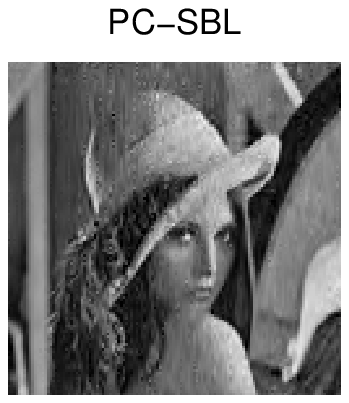}
& \hspace*{-10ex}
\includegraphics[width=5cm]{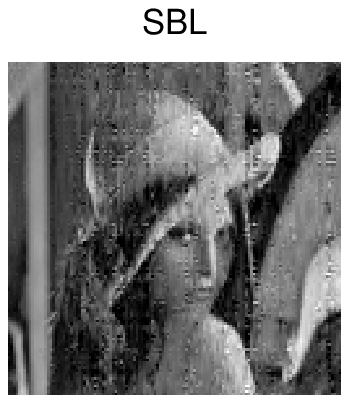}
& \hspace*{-10ex}
\includegraphics[width=5cm]{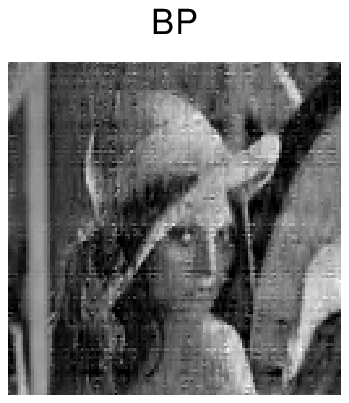}
\\
\hspace*{-3ex}\includegraphics[width=5cm]{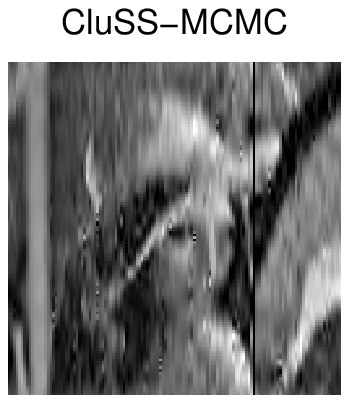} &
\hspace*{-10ex}
\includegraphics[width=5cm]{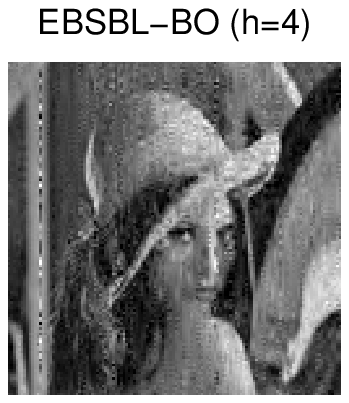}
& \hspace*{-10ex}
\includegraphics[width=5cm]{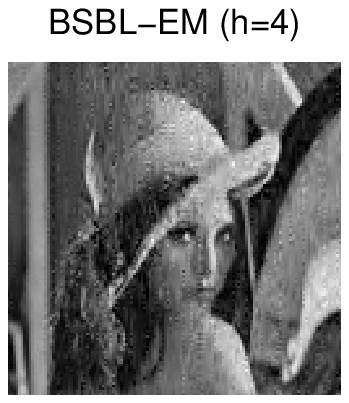}
& \hspace*{-10ex}
\includegraphics[width=5cm]{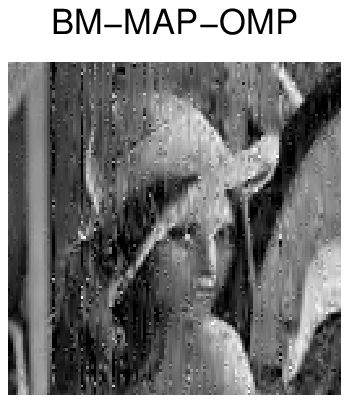}
\end{tabular}
  \caption{The original image `Lena' and the
reconstructed images using respective algorithms.}
   \label{fig4}
\end{figure*}

\begin{figure*}[!t]
 \centering
\begin{tabular}{cccccc}
\hspace*{-3ex}
\includegraphics[width=5cm]{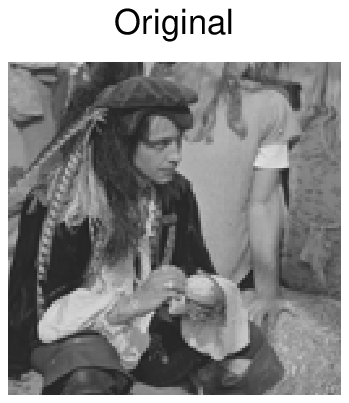}&
\hspace*{-10ex}
\includegraphics[width=5cm]{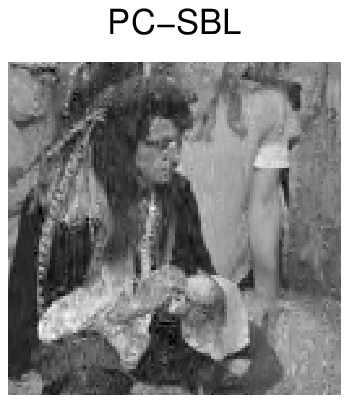}
& \hspace*{-10ex}
\includegraphics[width=5cm]{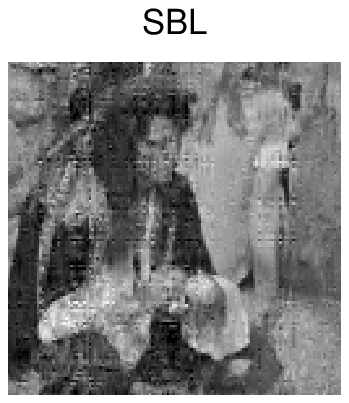}
& \hspace*{-10ex}
\includegraphics[width=5cm]{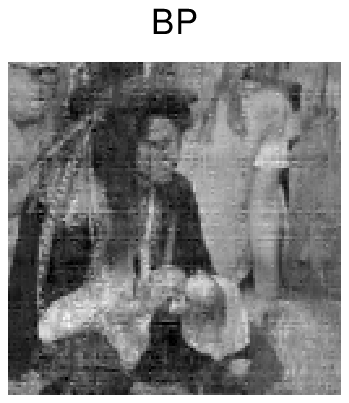}
\\
\hspace*{-3ex}\includegraphics[width=5cm]{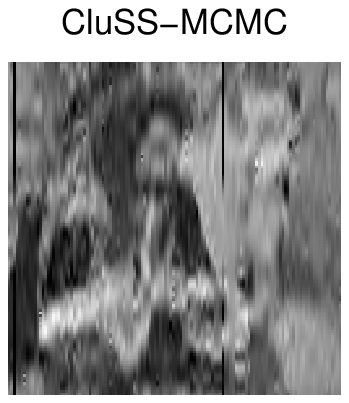} &
\hspace*{-10ex}
\includegraphics[width=5cm]{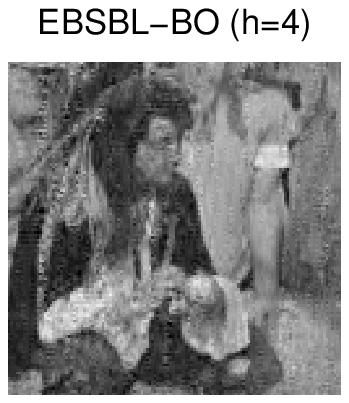}
& \hspace*{-10ex}
\includegraphics[width=5cm]{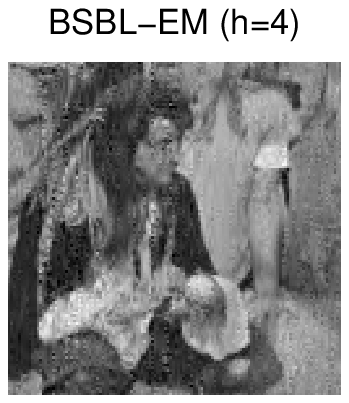}
& \hspace*{-10ex}
\includegraphics[width=5cm]{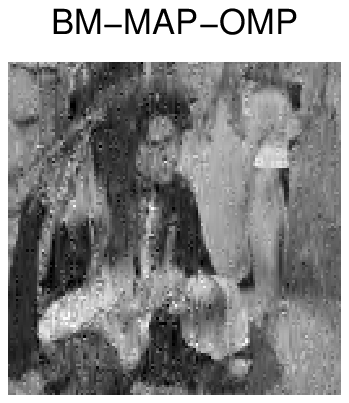}
\end{tabular}
  \caption{The original image `Pirate' and the
reconstructed images using respective algorithms.}
   \label{fig5}
\end{figure*}

\begin{figure*}[!t]
 \centering
\begin{tabular}{cccc}
\hspace*{-3ex}
\includegraphics[width=4.5cm]{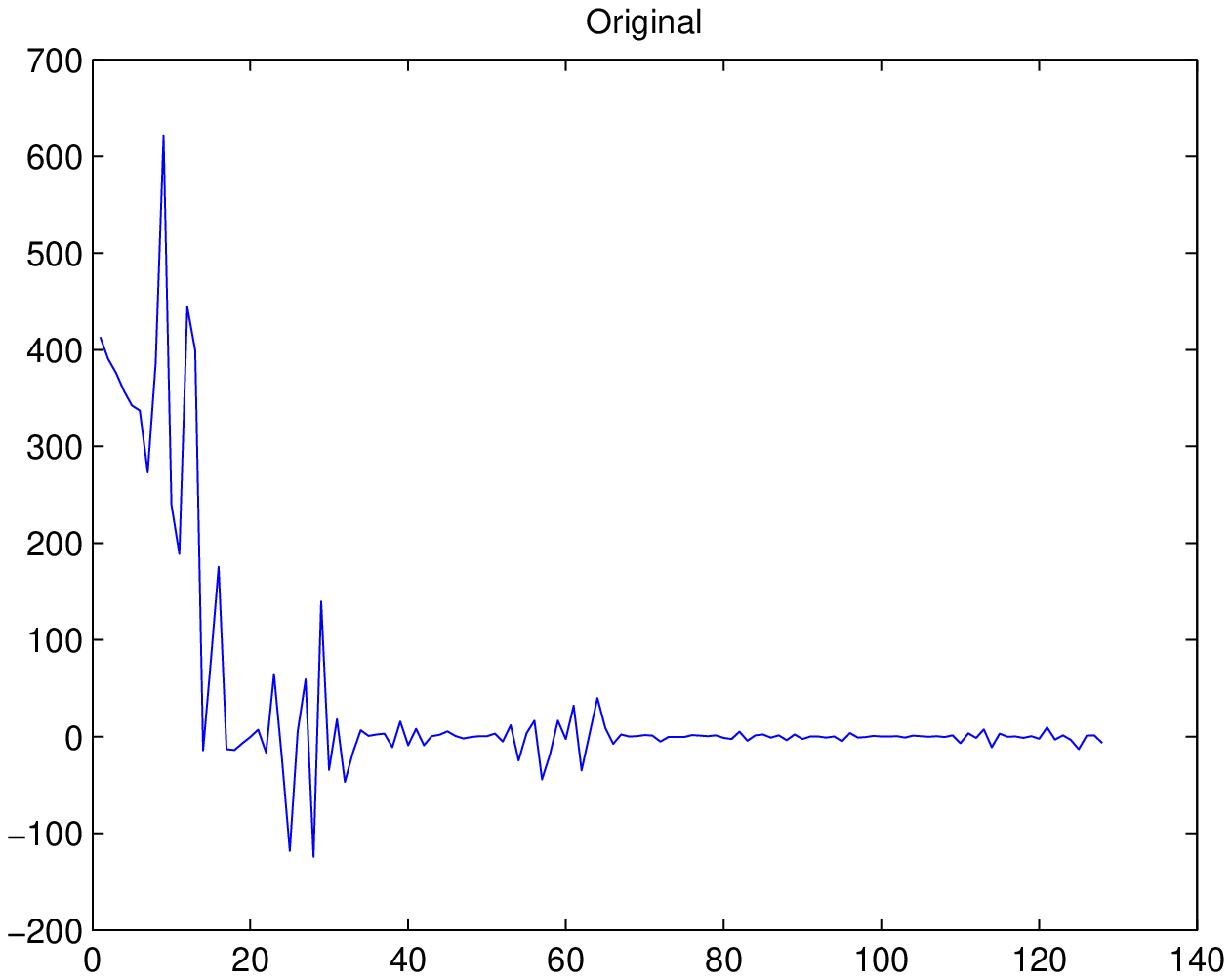}&
\hspace*{-5ex}
\includegraphics[width=4.5cm]{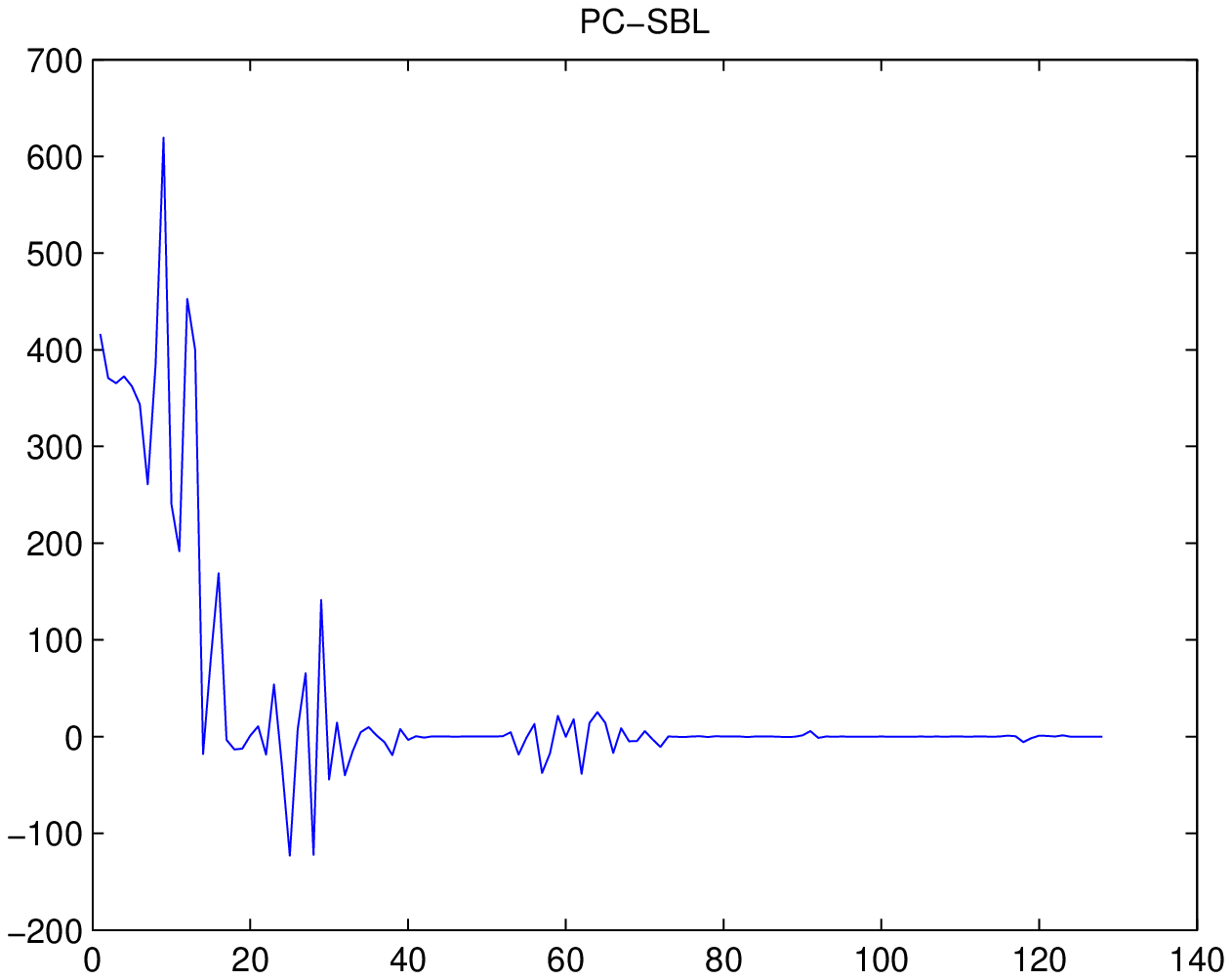}
& \hspace*{-5ex}
\includegraphics[width=4.5cm]{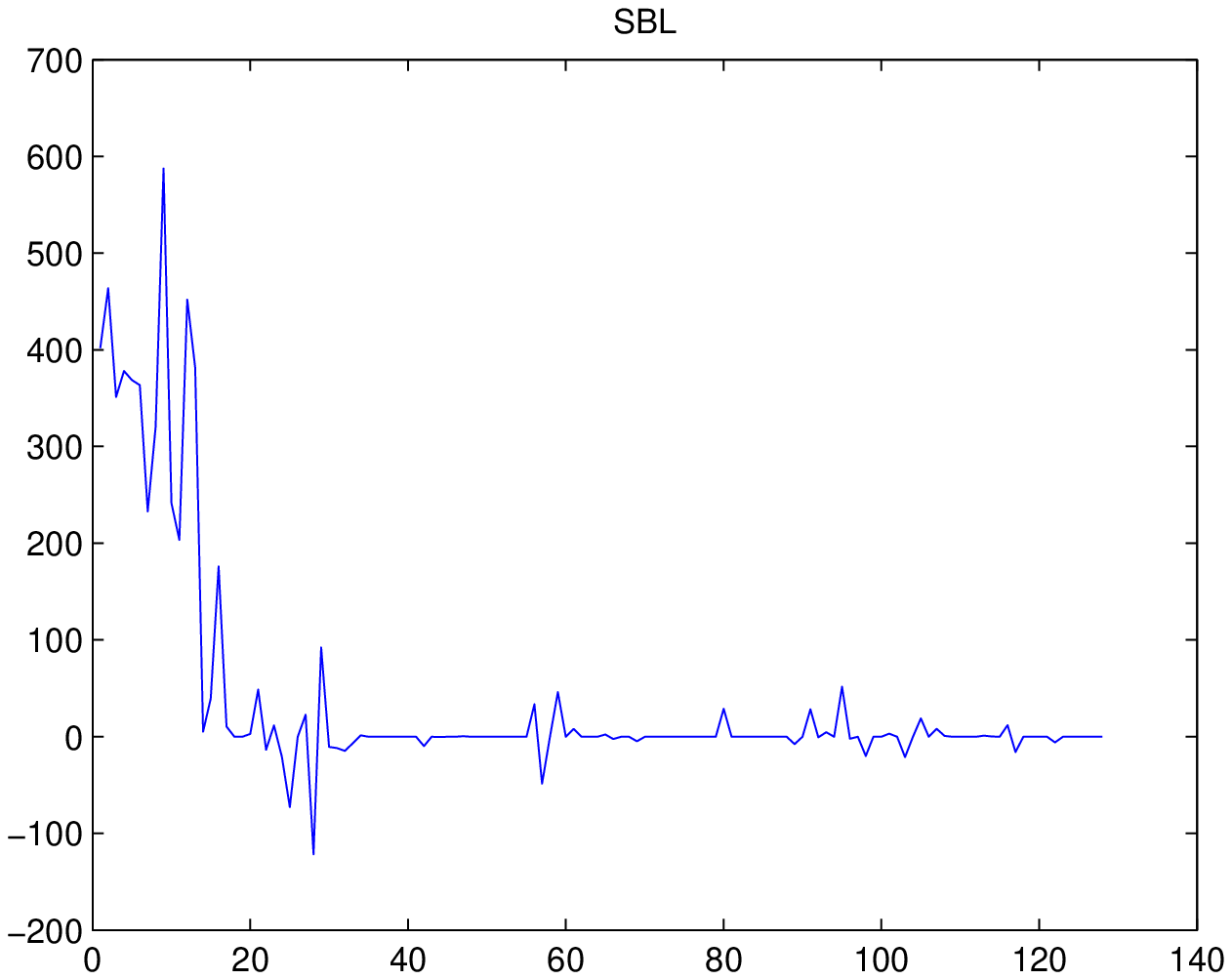}
& \hspace*{-5ex}
\includegraphics[width=4.5cm]{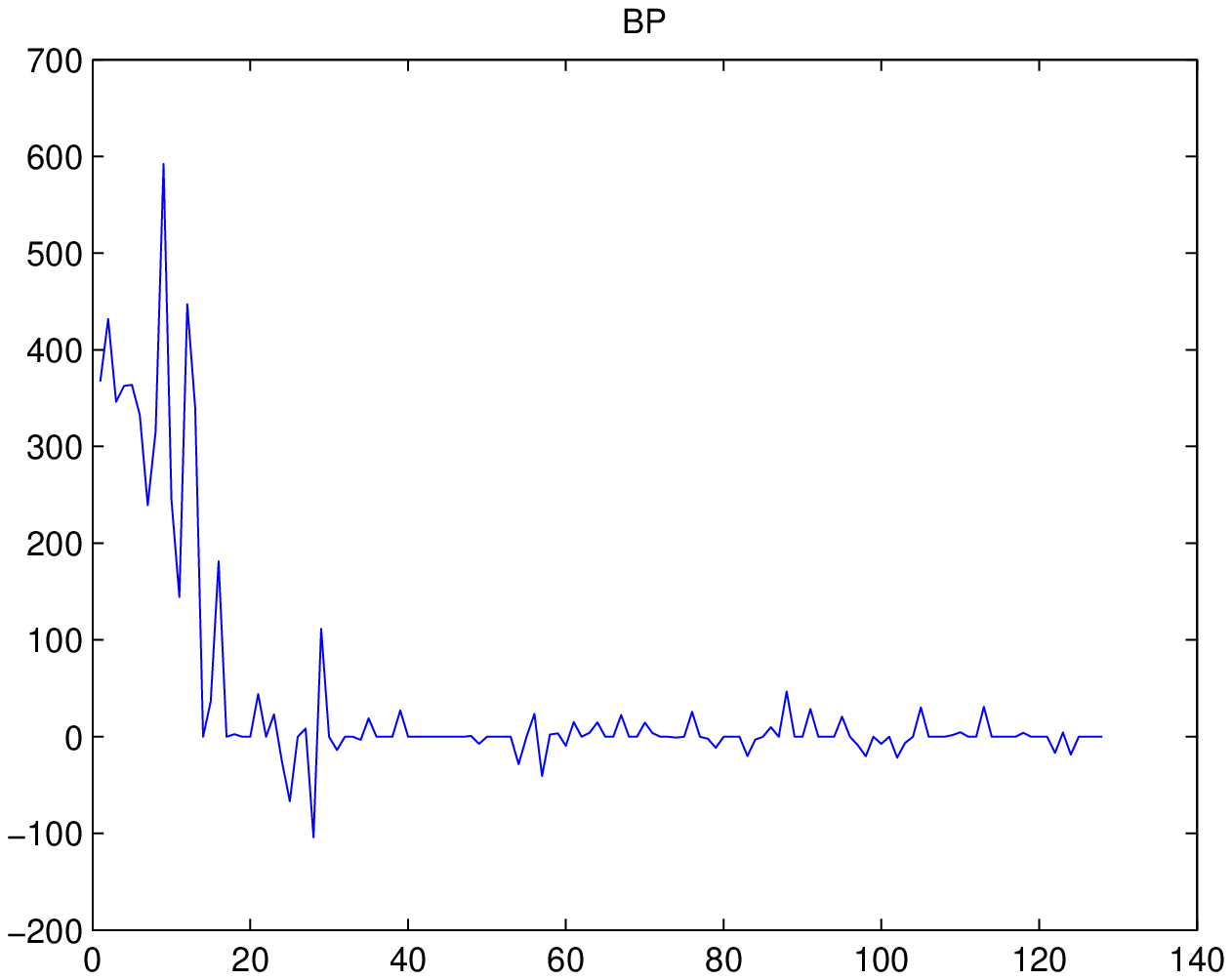}
\\
\hspace*{-3ex}\includegraphics[width=4.5cm]{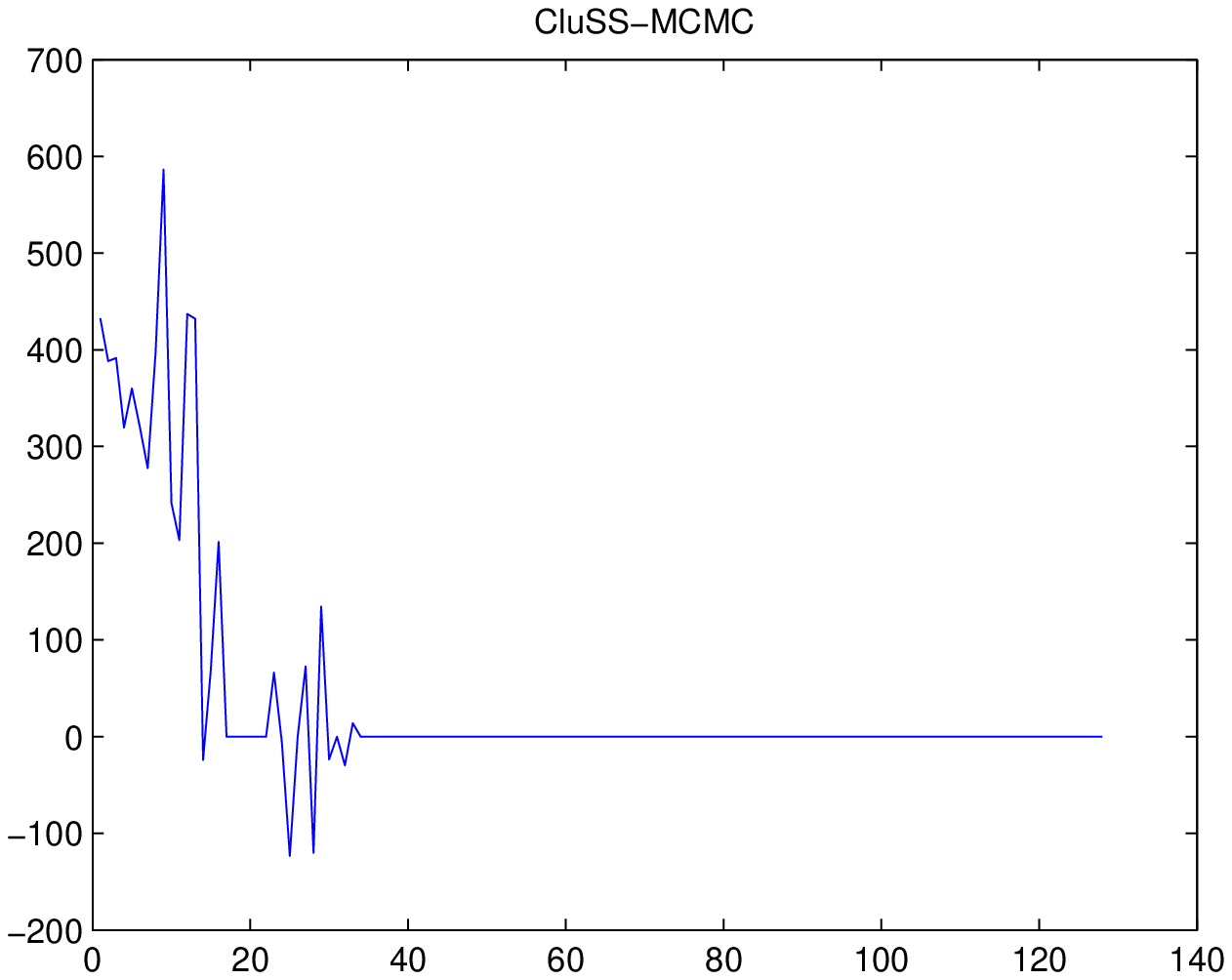} &
\hspace*{-5ex}
\includegraphics[width=4.5cm]{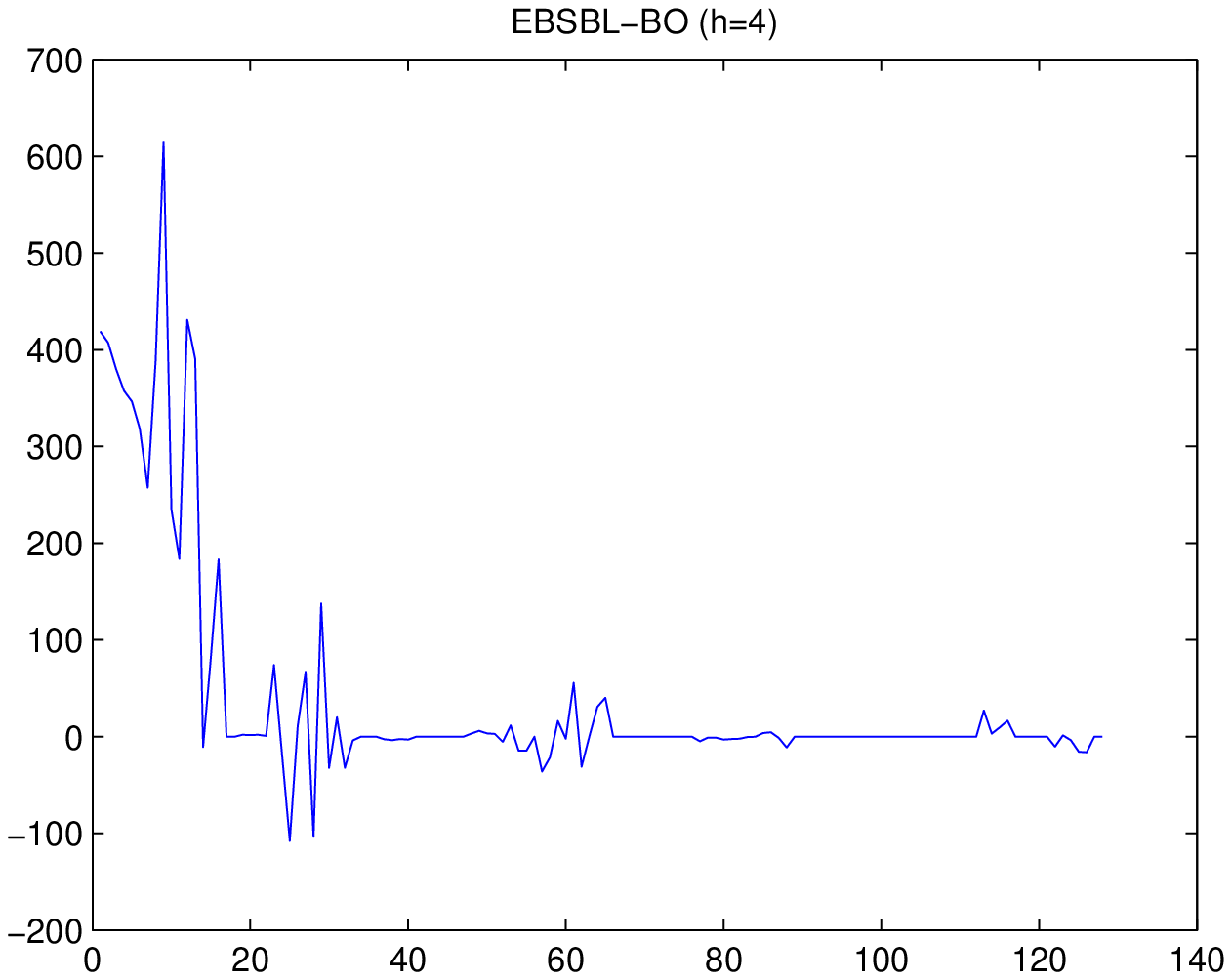}
& \hspace*{-5ex}
\includegraphics[width=4.5cm]{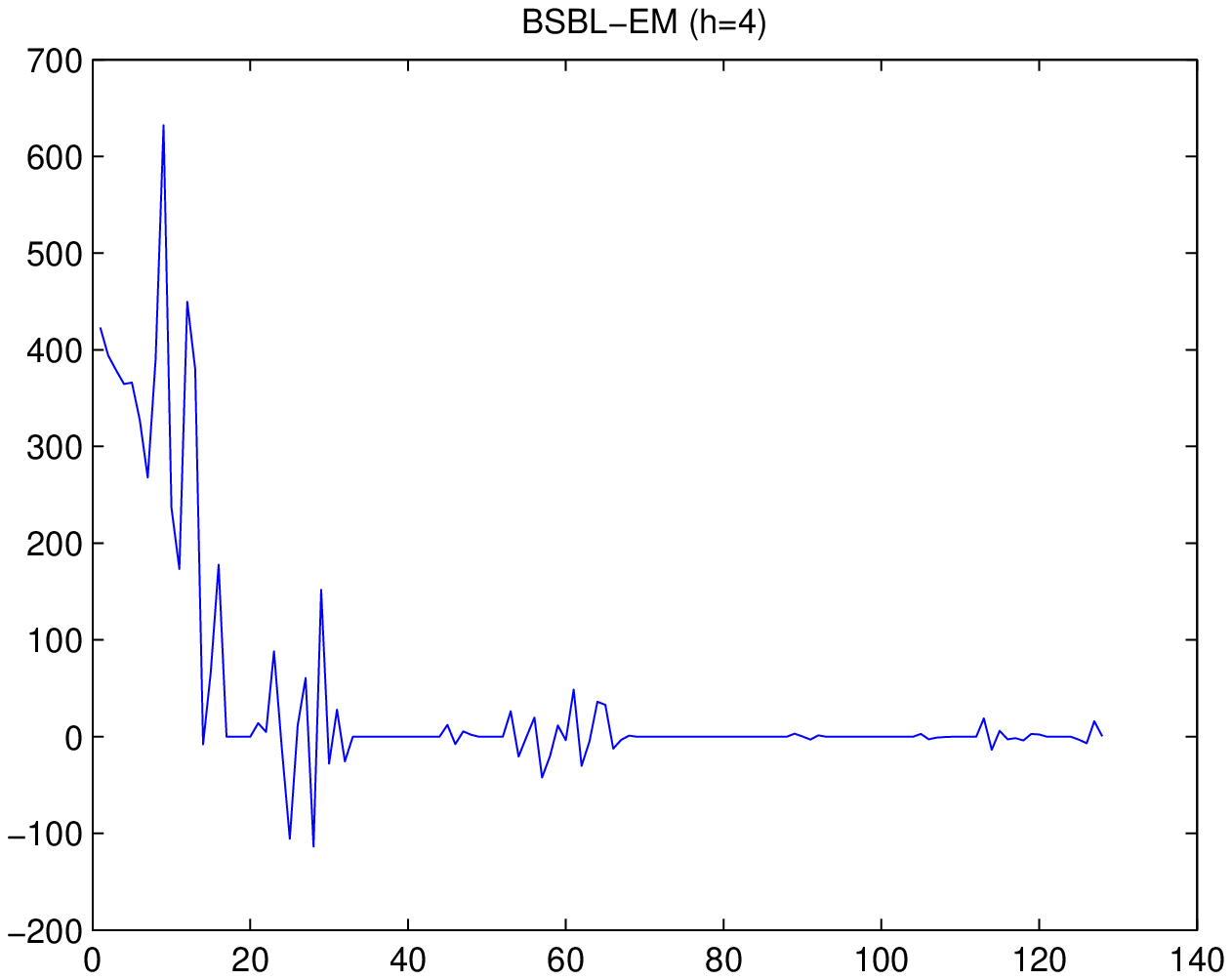}
& \hspace*{-5ex}
\includegraphics[width=4.5cm]{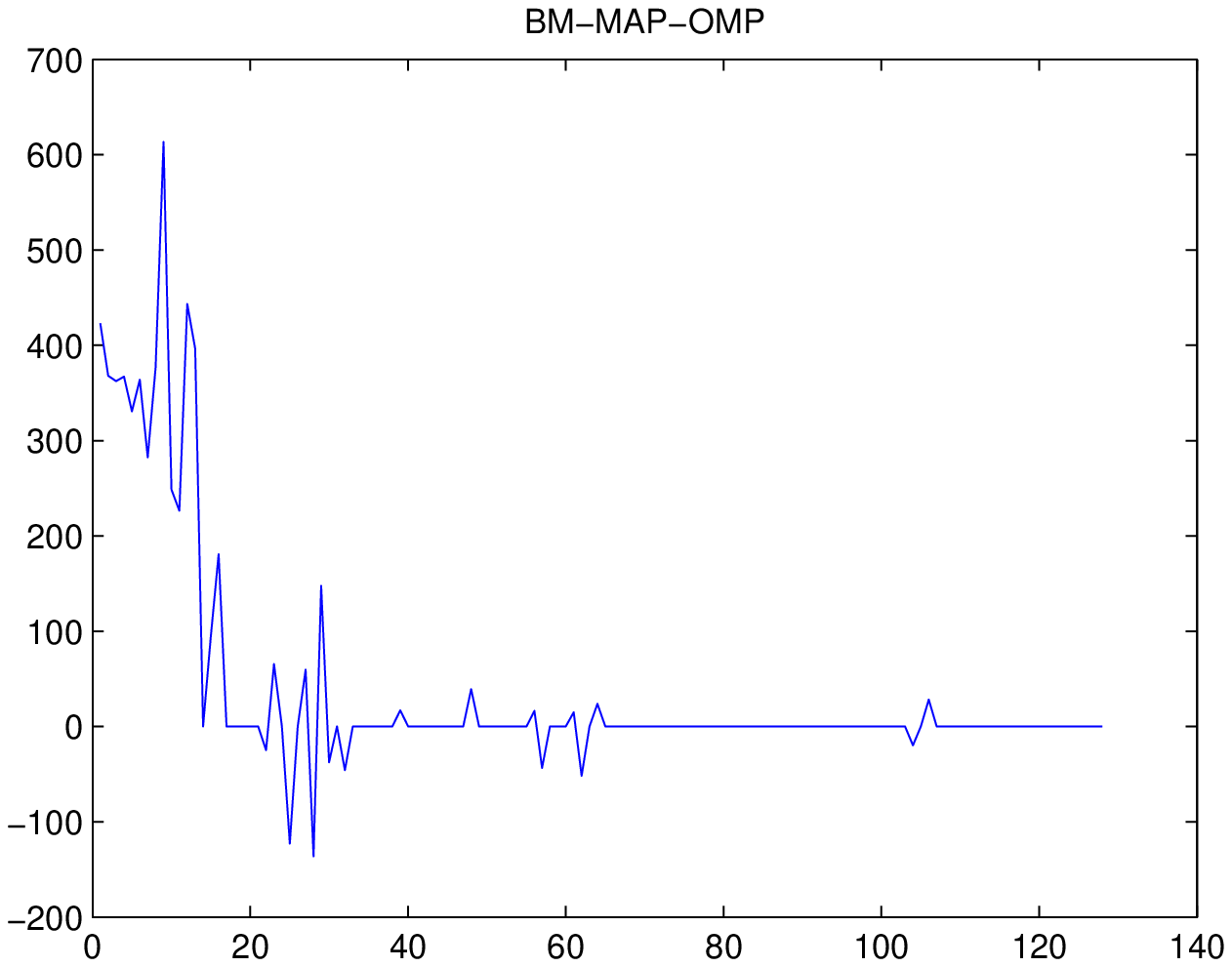}
\end{tabular}
  \caption{The true wavelet coefficients and the coefficients
reconstructed by respective algorithms.}
   \label{fig6}
\end{figure*}

In this subsection, we carry out experiments on real world images.
As it is well-known, images have sparse (or approximately sparse)
structures in certain over-complete basis, such as wavelet or
discrete cosine transform (DCT) basis. Moreover, the sparse
representations usually demonstrate clustered structures whose
significant coefficients tend to be located together (see Fig.
\ref{fig6}). Therefore images are suitable data sets for
evaluating the effectiveness of a variety of block-sparse signal
recovery algorithms. We consider two famous pictures `Lena' and
`Pirate' in our simulations. In our experiments, the image is
processed in a columnwise manner: we sample each column of the
$128\times 128$ image using a randomly generated measurement
matrix $\boldsymbol{A}\in \mathbb{R}^{m\times 128}$, recover each
column from the $m$ measurements, and reconstruct the image based
on the $128$ estimated columns. Fig. \ref{fig4} and \ref{fig5}
show the original images `Lena' and `Pirate' and the reconstructed
images using respective algorithms, where we set $m=64$ and $m=80$
respectively. We see that our proposed algorithm presents the
finest image quality among all methods. The result, again,
demonstrates its superiority over other existing methods. The
reconstruction accuracy of respective algorithms can also be
observed from the reconstructed wavelet coefficients. We provide
the true wavelet coefficients of one randomly selected column from
the image `Lena', and the wavelet coefficients reconstructed by
respective algorithms. Results are depicted in Fig. \ref{fig6}. It
can be seen that our proposed algorithm provides reconstructed
coefficients that are closest to the groundtruth.





\section{Conclusions} \label{sec:conclusion}
We developed a new Bayesian method for recovery of block-sparse
signals whose block-sparse structures are entirely unknown. A
pattern-coupled hierarchical Gaussian prior model was introduced
to characterize both the sparseness of the coefficients and the
statistical dependencies between neighboring coefficients of the
signal. The prior model, similar to the conventional sparse
Bayesian learning model, employs a set of hyperparameters to
control the sparsity of the signal coefficients. Nevertheless, in
our framework, the sparsity of each coefficient not only depends
on its corresponding hyperparameter, but also depends on the
neighboring hyperparameters. Such a prior has the potential to
encourage clustered patterns and suppress isolated coefficients
whose patterns are different from their respective neighbors. The
hyperparameters, along with the sparse signal, can be estimated by
maximizing their posterior probability via the
expectation-maximization (EM) algorithm. Numerical results show
that our proposed algorithm achieves a significant performance
improvement as compared with the conventional sparse Bayesian
learning method through exploiting the underlying block-sparse
structure, even without knowing the exact locations and sizes of
the non-zero blocks. It also demonstrates its superiority over
other existing methods and provides state-of-the-art performance
for block-sparse signal recovery.

\end{document}